\documentclass[epsfig,12pt,onecolumn]{article}
\usepackage[pdftex]{graphicx}
\usepackage{amsfonts}
\usepackage{amssymb}
\usepackage{multicol}
\usepackage{graphicx}
\usepackage{float}
\usepackage{caption}
\usepackage{xcolor}
\usepackage[utf8]{inputenc}
\usepackage{amsmath}

\newtheorem{theorem}{Theorem}

\newtheorem{conjecture}[theorem]{Conjecture}

\makeatletter \@addtoreset{equation}{section}

\def \be{\begin{equation}}
\def \ee{\end{equation}}
\def \bea{\begin{eqnarray}}
\def \eea{\end{eqnarray}}

\newcommand{\nc}{\newcommand}
\nc{\al}{\alpha} \nc{\bib}{\bibitem} \nc{\la}{\lambda}
\nc{\C}{\mbox{\hspace{1.24mm}\rule{0.2mm}{2.5mm}\hspace{-2.7mm} C}}
\nc{\R}{\mbox{\hspace{.04mm}\rule{0.2mm}{2.8mm}\hspace{-1.5mm} R}}

\begin{document}

\title{\textbf{Minimal Weak Gravity Conjecture and gauge duality in M-theory
on }$K3\times T^{2}$}
\author{M.Charkaoui$^{1,2}$, R. Sammani$^{1,2}$, E.H Saidi$^{1,2,3}$, R. Ahl
Laamara$^{1,2}$ \\
{\small 1. LPHE-MS, Science Faculty}, {\small Mohammed V University in
Rabat, Morocco}\\
{\small 2. Centre of Physics and Mathematics, CPM- Morocco}\\
{\small 3. Hassan II Academy of Science and Technology, Kingdom of Morocco}\\
{\small Email: mohammed\_charkaoui3@um5.ac.ma}}
\maketitle

\begin{abstract}
The minimal Weak Gravity Conjecture (WGC) predicts the emergence of towers
of superextremal states in both weak and strong coupling limits. In this
work, we study\textrm{\ }M-theory compactified on a special class of
Calabi-Yau threefolds to construct a 5D effective field theory (EFT) that
accommodates both weak and strong gauge coupling limits. Building on a
classification of fiber structures of Calabi-Yau threefolds with finite
volume, we establish a correspondence between curves in the fiber and the
base, which relates weak and strong gauge couplings. This allows us to probe
non-perturbative effects by treating strong couplings through their weakly
counterparts. We use this result and properties of
Bogomol'nyi--Prasad--Sommerfield (BPS) states to demonstrate that M-theory
on such Calabi-Yau threefold exhibits towers of superextremal BPS states in
the aforementioned extreme limits as expected by the minimal WGC.

\textbf{Keywords}: {\small minimal WGC, M-theory on CY3, \ Weak/Strong gauge
duality. }
\end{abstract}

\tableofcontents

\section{Introduction}

The tower versions \cite{1A,1B} of the Weak Gravity Conjecture (WGC) \cite%
{2A,2B,2C} (see also \cite%
{2E,2F} for reviews) were introduced to address certain inconsistencies of this
Swampland criterion \cite{3A,3B,3D} (see \cite{3E,3F,3H} for reviews on the Swampland Program) under circle reduction \cite{4A}.\
Various tests have been carried out to verify their validity; for instance,
the presence of towers of superextremal states in
compactifications of F-theory \cite{5,6,7} and M- theory \cite{5d,3d} to lower dimensional
effective field theories (EFT) in spacetime dimensions ranging from 6 to 3
has been thoroughly demonstrated. However, as pointed out in \cite{min},
there are cases where such towers have not been identified using current
techniques. It\ was then argued that towers of states should appear if and
only if they are required by the consistency of the WGC under circle
reduction.\ This is the case\ for\ Emergent string limits (weakly coupled),
Kaluza Klein reductions with KK gauge bosons and strong gauge coupling limit 
\cite{min}.

The weak gauge coupling limit has been explored in the context of M-theory
on Calabi-Yau threefolds (CY3) within the framework of the Asymptotic WGC 
\cite{5d}. In particular it was shown\ that, at infinite distances in the
Kahler moduli space, Calabi-Yau threefolds with finite volume exhibit a
special fibration structure classified as Type-$T^{2}$ or Type $K3/T^{4}$ 
\cite{8}. Moreover, towers of superextremal weakly coupled states emerge
from wrapping M2 and M5 branes on appropriate cycles in the fiber. A more
general investigation in \cite{9} further demonstrated that, in 5D EFT
descending from M-theory on Calabi-Yau threefolds, all
Bogomol'nyi--Prasad--Sommerfield (BPS) states arising from wrapping branes
on movable curves are indeed superextremal.

The analysis was further extended to M-theory compactified on a CY4\ in \cite%
{3d} where the finiteness of the fourfold's volume imposes constraints on
its structure, which can take the form $X_{4}=K3\times K3$, invoking a $%
Z_{2} $-automorphism relating weak and strong gauge coupling limits. The
resulting 3D EFT contains weakly and strongly coupled towers of
superextremal states. This setup provided not only a test for the Asymptotic
WGC but also an interesting realisation of the minimal WGC.\ Recall that
this later conjecture is a refined version of the WGC proposed to solve the
aforementioned inconsistencies under circle reduction. This refined version
postulates the existence of towers of superextremal particle states $%
\{m_{k}\}$\ below the black hole threshold $M_{\mathrm{BH},\min },$\ if and
only if at least one of the following statements holds: There is an emergent
string limit, a reduction with KK gauge bosons, or a strong coupling limit 
\cite{min}.\

In this paper, we test the minimal WGC for the case of 5D EFT descending
from M-theory on CY3 focusing on a unique case within its possible fibration
structures along the lines of \cite{8}. By analysing the geometry of the
manifold, we establish a correspondence between weak and strong gauge
couplings.\ The existence of these limits provides the first indication
towards the validity of the Minimal WGC. After constructing candidate towers
of states by wrapping branes on appropriate cycles of the manifold, we show
that these states are indeed superextremal, confirming that the conjecture
is satisfied.

To establish the structure of the manifold, we use the fact that in the
infinite distance limit of the moduli space, formally designated by the
spectral parameter $\lambda \rightarrow \infty $, the fiber shrinks while
the base expands such that the manifold exhibits either a torus $T^{2}$
fibration, or a surface $S=K3,T^{4}$ fibration \cite{8}. The novelty of our
approach lies in allowing both the fiber and the base to shrink and expand,
which corresponds to taking the limits $\lambda \rightarrow 0,\infty $. This
introduces a constraint on the base as well, which should also be either $%
T^{2},$ $K3$ or $T^{4}$. We deduce that the manifold takes the form $%
X_{3}=K3\times T^{2}$ or $T^{4}\times T^{2}.$ The structure of the internal
manifold induces weakly and strongly coupled directions in the charge
lattice, depending on whether these directions arise from shrinking or
expanding cycles in the threefold. We prove that there exists a weak/strong
gauge duality that is generated by the mapping $\lambda \leftrightarrow
1/\lambda $, relating the volumes of the curves in the fiber and the base of
the manifold ensuring that the full volume of the CY3 throughout the moduli
space remains finite.

This duality aligns with the results obtained in \cite{3d} where both the
fiber and the base are $K3$ surfaces. By imposing these conditions on the
volumes of the fiber, base and total manifold, we find as mentioned in \cite%
{min}, that instead of a circle reduction from 6 to 5 dimensions, using
F-/M-theory duality, the 5D theory can be viewed as either an emergent
string limit or a decompactification limit. And as expected by the minimal
WGC, weakly and strongly coupled towers of superextremal states are indeed
present; they emerge from M2 and M5 branes wrapping fiber or base cycles of
the CY3.

The structure of this paper is as follows: In the second section, we first
review the fibration structure of finite volume Calabi-Yau threefolds at
infinite distances. Then we give our first result regarding the extended
limits (both $\lambda \rightarrow 0,\infty $) and the associated geometrical
implications.\ In the third section, we show that this seemingly purely
geometric correspondence between the fiber and the base given by the mapping 
$\lambda \rightarrow 1/\lambda $ relates also weak and strong gauge
couplings.\ Finally we examine the towers of states across these distinct
geometries and gauge regimes then we identify the BPS towers of states
satisfying the minimal WGC.

\section{Calabi-Yau threefolds in extreme limits}

In this section, we focus on constructing 5D EFTs from M-theory compactified
on Calabi-Yau threefolds with the goal of investigating extreme gauge
coupling limits (weak and strong) by appropriately architecting the internal
manifold. First, we review Calabi-Yau threefold fibrations in infinite
distance limits, parameterized by a spectral parameter $\lambda \rightarrow
\infty $\ \cite{8, 5d} to identify possible structures that can give rise to
a finite volume. Then, we probe the previously underinvestigated limit $%
\lambda \rightarrow 0$\ using the approach of \cite{3d} and establish a
correspondence between curves in the fiber and those in the base for
different manifold configurations.

\subsection{Review and preliminary results}

Following \cite{8, 5d}, Calabi-Yau threefolds $X_{3}$ with finite volume $%
\mathcal{V}_{X_{3}}$ admit two types of fibrations distinguished by the
dimension of the fiber $\mathcal{F}$ and, correspondingly, the base $%
\mathcal{B}$. Particularly:

\begin{description}
\item[$\left( 1\right) $] Type -$\mathbb{T}^{2}$ fibration: the fiber $%
\mathcal{F}_{1}$ is a 2-torus (a complex line), and the base $\mathcal{B}%
_{2} $ is a two complex dimensional geometry that can be taken as one of the
Hirzebruch surfaces $F_{n}$ \cite{9A}$.$

\item[$\left( 2\right) $] Type-$\mathbb{S}$ fibration: the fiber $\mathcal{F}%
_{2}$ here is a complex surface $\mathbb{S}$, which can be either a K3
surface or a 4-torus $\mathbb{T}^{4}$ \cite{3d, 5d}. For this type of
fibration, the complex base $\mathcal{B}_{1}$ is one-dimensional. Examples
of $\mathcal{B}_{1}$ include the complex projective line $\mathbb{P}^{1},$
isomorphic to the real 2-sphere.
\end{description}

Accordingly, Calabi-Yau threefolds $X_{3}$ can be classified into two main
types: Type -$\mathbb{T}^{2}$ and Type-$\mathbb{S}$. They can be formally
expressed as follows: 
\begin{equation}
X_{3}\sim \mathcal{F}_{n}\times \mathcal{B}_{3-n}\qquad ,\qquad n=1,2
\label{X3}
\end{equation}%
Generic curves $C$ in the CY3 split into $C_{F}+C_{B};$ each component is
given by linear combinations of 2-cycles in the fiber $\mathcal{F}_{n}$ and
the base $\mathcal{B}_{3-n}$ as follows:%
\begin{equation}
C=\sum_{\alpha \in fiber}\tilde{q}_{\alpha }\tilde{C}^{\alpha }+\sum_{a\in
base}q_{a}C^{a}  \label{C}
\end{equation}%
where $Q_{A}=(\tilde{q}_{\alpha },q_{a})$ represent the integer charges of
the multi $U(1)$ symmetries resulting from M-theory compactification on the
CY3. We also consider the total volume of the CY3 as the product of the
volumes of the fiber and the base like $\mathcal{V}_{X_{3}}\sim \mathcal{V}_{%
\mathcal{F}_{n}}\times \mathcal{V}_{\mathcal{B}_{3-n}}$. Under the
finiteness condition $\mathcal{V}_{X_{3}}=\mathrm{cte}$, the product
translates into a relationship between $\mathcal{V}_{\mathcal{F}_{n}}$ and $%
\mathcal{V}_{\mathcal{B}_{3-n}}$ as: 
\begin{equation}
\mathcal{V}_{\mathcal{F}_{n}}\times \mathcal{V}_{\mathcal{B}_{3-n}}=\mathrm{%
cte}\qquad \Leftrightarrow \qquad \mathcal{V}_{\mathcal{B}_{3-n}}=\frac{%
\mathrm{cte}}{\mathcal{V}_{\mathcal{F}_{n}}}
\end{equation}

\begin{figure}[h]
\begin{center}
\includegraphics[width=10cm]{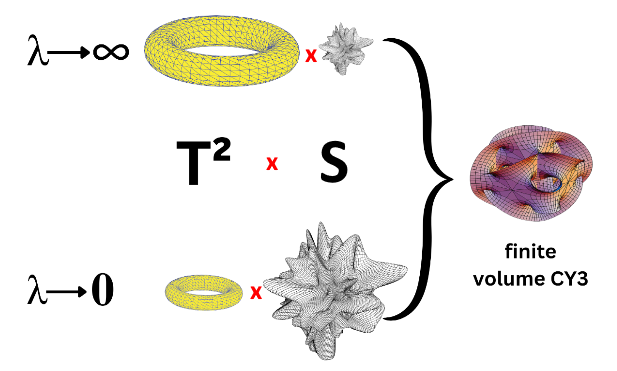}
\end{center}
\caption{Two extreme configurations of the fibrations of the Calabi-Yau
threefolds according to the values of the spectral parameter. They are
labeled by the limits $\protect\lambda \rightarrow \infty $ \ and $\protect%
\lambda \rightarrow 0$.}
\label{figTS}
\end{figure}

In what follows, we study both classes of fibrations (\ref{X3}) in
particular configurations corresponding to the singular limits including the
usual 
\begin{equation}
\begin{tabular}{|c|c|c|c|}
\hline
configutation & fiber & base & CY3 \\ \hline
$\left( i\right) $ & $\mathcal{V}_{\mathcal{F}_{1}}\rightarrow 0$ & $%
\mathcal{V}_{\mathcal{B}_{2}}\rightarrow \infty $ & $\mathcal{V}_{\mathcal{F}%
_{1}}\mathcal{V}_{\mathcal{B}_{2}}=\mathrm{cte}$ \\ \hline
$\left( ii\right) $ & $\mathcal{V}_{\mathcal{F}_{2}}\rightarrow 0$ & $%
\mathcal{V}_{\mathcal{B}_{1}}\rightarrow \infty $ & $\mathcal{V}_{\mathcal{F}%
_{2}}\mathcal{V}_{\mathcal{B}_{1}}=\mathrm{cte}$ \\ \hline
\end{tabular}
\label{5}
\end{equation}%
and our extension:%
\begin{equation}
\begin{tabular}{|c|c|c|c|}
\hline
configutation & fiber & base & CY3 \\ \hline
$\left( iii\right) $ & $\mathcal{V}_{\mathcal{F}_{1}}\rightarrow \infty $ & $%
\mathcal{V}_{\mathcal{B}_{2}}\rightarrow 0$ & $\mathcal{V}_{\mathcal{F}_{1}}%
\mathcal{V}_{\mathcal{B}_{2}}=\mathrm{cte}$ \\ \hline
$\left( iv\right) $ & $\mathcal{V}_{\mathcal{F}_{2}}\rightarrow \infty $ & $%
\mathcal{V}_{\mathcal{B}_{1}}\rightarrow 0$ & $\mathcal{V}_{\mathcal{F}_{2}}%
\mathcal{V}_{\mathcal{B}_{1}}=\mathrm{cte}$ \\ \hline
\end{tabular}
\label{54}
\end{equation}%
An illustration of these extreme limits for the Type-$\mathbb{T}^{2}$
fibrations is shown in the Figure \ref{figTS} 
To conduct this study, we borrow ideas from \cite{8} where one parameterises
the volumes of the fiber $\mathcal{F}_{n}$ and the base $\mathcal{B}_{3-n}$
by a spectral parameter $\lambda $ like%
\begin{equation}
\mathcal{V}_{\mathcal{F}_{n}}=F(\lambda )\qquad ,\qquad \mathcal{V}_{%
\mathcal{B}_{3-n}}=B(\lambda )
\end{equation}%
such that $F(\lambda )B(\lambda )=\mathrm{cte}$, indicating that the total
volume $\mathcal{V}_{X_{3}}$ is independent of $\lambda $. Moreover, in \cite%
{5d}, the configurations (\ref{5}) have been imagined in terms of the
asymptotic limit $\lambda \rightarrow \infty $ as,%
\begin{equation}
\lim_{\lambda \rightarrow \infty }\mathcal{V}_{\mathcal{F}_{n}}\rightarrow
0\qquad ,\qquad \lim_{\lambda \rightarrow \infty }\mathcal{V}_{\mathcal{B}%
_{3-n}}\rightarrow \infty  \label{6}
\end{equation}%
and have been given an interesting physical interpretation in link with
gauge coupling regimes \cite{5d}. More concretely, the associated Kahler
form of the Calabi-Yau $X_{3}$ satisfying (\ref{5}) can be written down as
follows:%
\begin{equation}
J=\lambda ^{\alpha }\mathcal{J}_{b}+\frac{1}{\lambda ^{\beta }}\mathcal{J}%
_{f}
\end{equation}%
where $\mathcal{J}_{b}$ and $\mathcal{J}_{f}$ respectively represent the
Kahler form generators of the base $\mathcal{B}$ and the fiber $\mathcal{F}$
with positive $\alpha \times \beta $. And given that the total volume of the
Calabi-Yau can be formulated as:%
\begin{equation}
\mathcal{V}_{X_{3}}\mathcal{=}\frac{1}{6}\int_{X_{3}}J^{3}\qquad \mathcal{V}%
_{T^{2}}\mathcal{=}\int_{T^{2}}J\qquad \mathcal{V}_{B}\mathcal{=}\frac{1}{2}%
\int_{\mathbb{S}}J^{2}
\end{equation}%
it is clear that at the infinite distance limit $\lambda \rightarrow \infty
, $ the condition on the finiteness of the volume explicitly defines the two
classes of the fibrations\textrm{\ }Type -$\mathbb{T}^{2}$ and Type-$\mathbb{%
S}$ as follows:\textrm{\ }

\begin{itemize}
\item Type -$\mathbb{T}^{2}:$ $\mathcal{J}_{b}^{3}=0,$ $\mathcal{J}%
_{b}^{2}\neq 0$ with Kahler form given by:%
\begin{equation}
J=(\lambda )^{\frac{1}{3}}\mathcal{J}_{b}+(\frac{1}{\lambda })^{\frac{2}{3}}%
\mathcal{J}_{f}  \label{T2}
\end{equation}

The vanishing condition on the triple intersection of a nef divisor $%
\mathcal{J}_{b}^{3}=0$, in addition to the non vanishing intersection of the
two divisors $\mathcal{J}_{b}^{2}\neq 0$ is known as the Oguiso criterion;
it characterises the $T^{2}$ fibration of Calabi-Yau threefold $X_{3}$. Such
Calabi-Yaus were constructed in \cite{8} where the fiber and the base can be
shown to scale as $\mathcal{V}_{T^{2}}\sim \lambda ^{-\frac{2}{3}},\mathcal{V%
}_{\mathbb{B}_{2}}\sim \lambda ^{\frac{2}{3}},$ with the scaling exponent $%
\nu =\frac{2}{3}$ maintaining the overall volume finite $\mathcal{V}%
_{X_{3}}= $cte.

\item Type-$\mathbb{S}:$ $\mathcal{J}_{b}^{3}=0,\ \mathcal{J}_{b}^{2}=0,$
the Kahler form is:%
\begin{equation}
J=(\lambda )^{\frac{2}{3}}\mathcal{J}_{b}+(\frac{1}{\lambda })^{\frac{1}{3}}%
\mathcal{J}_{f}  \label{K3}
\end{equation}%
The Calabi-Yau has either a $K3$ or $T^{4}$ fiber depending on the second
Chern class\textrm{. }The volumes of the fiber and the base behave as $%
\mathcal{V}_{K3/T^{4}}\sim \lambda ^{-\frac{2}{3}},\mathcal{V}_{\mathbb{B}%
_{1}}\sim \lambda ^{\frac{2}{3}}$.
\end{itemize}

A natural extension to this above classification is to consider, in addition
to the limit $\lambda \rightarrow \infty $, the regime $\lambda \rightarrow
0 $ and investigate the arising CY structure. By doing so, we show in the
next subsection that the base should also share the same structure, $T^{2}$
or $\mathbb{S},$ as the fiber i,e the manifold should take the form $%
T^{2}\times \mathbb{S}.$ This is unlike previous approaches which only fix
one extreme limit that constrains solely the fiber without imposing any
conditions on the structure of the base \cite{8, 5d}. The key distinction
between our study and previous works is summarized in the table below:\ 
\begin{equation*}
\begin{array}{ccc}
\text{values of }\lambda \text{ studied} & F\text{ fixed} & B \\ 
\lambda \rightarrow \infty \text{ \cite{5d}} & \text{Type-}\mathbb{T}^{2}%
\text{ or Type-}\mathbb{S} & \text{Free} \\ 
\lambda \rightarrow 0,\infty & \text{Type-}\mathbb{T}^{2}\text{ or Type-}%
\mathbb{S} & \text{fixed: Type-}\mathbb{T}^{2}\text{ or Type-}\mathbb{S}%
\text{ compatible with }F%
\end{array}%
\end{equation*}%
At last, we mention that this investigation parallels the analysis of \cite%
{3d} on Calabi-Yau fourfolds, where both the fiber and the base are allowed
to contract or expand by considering both limits: $\lambda \rightarrow
\infty $ and $\lambda \rightarrow 0$ while the total volume of the internal
manifold remains finite.

\subsection{Fixing the base of the fibered CYs}

Here, we explore the extreme limit $\lambda \rightarrow 0$ and study the
implications on the structure of the CY manifold in contrast with the
commonly studied case $\lambda \rightarrow \infty .$ Since fixing the latter
gives the fibration of the CY, the limit $\lambda \rightarrow 0$ will fix
the base as follows:

\begin{itemize}
\item Type-$\mathbb{T}^{2}$: in the limit $\lambda \rightarrow 0,$ the fiber 
$T^{2}$ expands as$\ $in (\ref{T2}). In order for the volume of CY to remain
finite, the following condition is required:%
\begin{equation}
\mathcal{J}_{f}^{3}=0\qquad ,\qquad \mathcal{J}_{f}^{2}=0
\end{equation}%
Correspondingly, the Calabi-Yau threefold admits a transverse component to $%
T^{2},$ a base, which is either a $K3$ or $T^{4}$ surface$:$%
\begin{equation}
\begin{tabular}{ccc}
$T^{2}$ & $\rightarrow $ & $X_{3}$ \\ 
&  & $\downarrow $ \\ 
&  & $K3$%
\end{tabular}%
\qquad ,\qquad 
\begin{tabular}{ccc}
$T^{2}$ & $\rightarrow $ & $X_{3}$ \\ 
&  & $\downarrow $ \\ 
&  & $T^{4}$%
\end{tabular}%
\end{equation}

\item Type-$\mathbb{S}$: again by taking the limit $\lambda \rightarrow 0$
in (\ref{K3})$,$ the surface fiber $K3$/$T^{4}$ expansion implies the
constraint:%
\begin{equation}
\mathcal{J}_{f}^{3}=0\qquad ,\qquad \mathcal{J}_{f}^{2}\neq 0
\end{equation}%
this gives rise to a base having the structure of a $T^{2}$-torus:%
\begin{equation}
\begin{tabular}{ccc}
$K3$ & $\rightarrow $ & $X_{3}$ \\ 
&  & $\downarrow $ \\ 
&  & $T^{2}$%
\end{tabular}%
\qquad ,\qquad 
\begin{tabular}{ccc}
$T^{4}$ & $\rightarrow $ & $X_{3}$ \\ 
&  & $\downarrow $ \\ 
&  & $T^{2}$%
\end{tabular}%
\end{equation}
\end{itemize}

To sum up, the structure of the Calabi-Yau threefold can be architectured in
one of the following ways: $X_{3}=F\times B$ and $X_{3}^{\prime }=F^{\prime
}\times B^{\prime }$ where%
\begin{eqnarray}
(F,B) &=&(T^{2},K3)\qquad ,\qquad (K3,T^{2})  \label{CY} \\
(F^{\prime },B^{\prime }) &=&(T^{2},T^{4})\qquad ,\qquad (T^{4},T^{2})
\end{eqnarray}%
with $X_{3}^{\prime }$ seen as a trivial Calabi-Yau threefold \cite{3E}. In
each of these cases, we can chose to treat either part of the CY3 as the
fiber or the base.

In what follows, we will focus on the case where the surface is $K3.$ Notice
that the mapping $\lambda \rightarrow 1/\lambda $ swaps (\ref{T2}) with (\ref%
{K3}), thus exchanging the roles of curves in the fiber with those of the
base.\ This correspondence is reminiscent of the $\mathbb{Z}_{2}$
automorphism of the Calabi-Yau fourfold $K3\times K3$ studied in \cite{3d}.\
In fact, one can replicate the same analysis to generalise this
correspondance\ for different fourfolds of the form $T^{2}\times CY3$\textrm{%
.}

With these preliminary results at hand, the next step is to identify the
different gauge coupling limits of our EFT.\ In the next section, we
investigate some of the implications of the compactification of M-theory on
one of the Calabi-Yau threefolds possessing the aforementioned properties.\
We take the example of $K3\times T^{2}$ and we examine the physical
implication for the 5D EFT. We show that the correspondence between curves
in the fiber and the base given by the mapping $\lambda \rightarrow
1/\lambda $ relates also weak and strong gauge couplings.\ The emergence of
these two gauge regimes leads us to examine the validity of the minimal WGC
proposed in \cite{min}. This refined version of the WGC predicts the
presence of towers of superextremal particles below the black hole threshold
in emergent strings \cite{8}, KK reduction, or strong gauge coupling limit.
A detailed analysis of the minimal WGC and its validity in our setting is
conducted in section 4.

\section{M-theory on $K3\times T^{2}$ and the weak/strong gauge regimes}

From the previous section we have deduced that constraining the internal
manifold to have a finite volume in both extreme limits leads to fixing the
geometrical structure of the fiber and base.\ This results in a Calabi-Yau
of the form $\mathbb{S}\times T^{2},$ where $\mathbb{S}$ is either an
abelian Schoen manifold or a K3 surface.\ In this section we aim to derive
the implications of the algebraic and topological structures of the fiber
and the base in the EFT.

We start by compactifying M-theory on Calabi-Yau threefolds $X_{3}$ of the
established form:%
\begin{equation}
\text{Case I}:%
\begin{tabular}{ccc}
$T^{2}$ & $\rightarrow $ & $X_{3}$ \\ 
&  & $\downarrow $ \\ 
&  & $K3$%
\end{tabular}%
\quad \text{,}\quad \text{Case II : }%
\begin{tabular}{ccc}
$K3$ & $\rightarrow $ & $X_{3}$ \\ 
&  & $\downarrow $ \\ 
&  & $T^{2}$%
\end{tabular}
\label{FIB}
\end{equation}%
This theory has been thoroughly studied and many of its features have been
examined and listed \cite{9B, 9C}.\emph{\ }In \cite{5d}, it was shown that a
special class $\mathcal{C}_{f}$ of curves in the fiber give rise to a weak
coupling limit when $\mathcal{V}_{\mathcal{C}_{f}}\rightarrow 0$.\ This
induces a diverging curve $C_{b}$\ in the base giving rise to a strong
coupling limit.\ This has been demonstrated in a simpler setup in the case
of a fourfold $K3\times K3$ \cite{3d} where the link between the fiber and
the base is more noticeable; here, we extend it to the $\mathbb{S}\times
T^{2}$ scenario at hand.

Let us begin by splitting the basis of curves of $X_{3}$ as:%
\begin{equation}
\{\mathcal{C}^{A}\}=\{\tilde{C}^{\alpha },C^{a}\}  \label{CA}
\end{equation}%
where the $\tilde{C}^{\alpha }$ denote the fibral curves and $C^{a}$
designate the base curves.\ For a generic curve $\mathcal{C}=q_{A}\mathcal{C}%
^{A}$ in the charge lattice $H_{2}(X_{3}),$ the gauge coupling $g_{\mathcal{C%
}}^{2}$ of the associated U(1)$_{\mathcal{C}}$ EFT is is given by $%
g_{5}^{2}(q_{A}G^{AB}q_{B}).$Here $g_{5}$ is the YM gauge coupling of the
compactified 5D gauge theory and $G^{AB}$ is the inverse of the intersection
matrix $G_{AB}$ (see appendix). By using (\ref{CA}) we get 
\begin{equation}
g_{\mathcal{C}}^{2}=g_{5}^{2}(q_{a}G^{ab}q_{b}+2q_{a}G^{^{\prime }a\alpha }%
\tilde{q}_{\alpha }+\tilde{q}_{\alpha }\tilde{G}^{\alpha \beta }\tilde{q}%
_{\beta })  \label{YM}
\end{equation}%
where $G^{ab}$ (resp.\ $\tilde{G}^{\alpha \beta }$) contains only
contributions from the base (resp.\ the fiber) and $G^{^{\prime }a\alpha }$
gets contributions from both. For $T^{2}\times K3$ in (\ref{FIB}-I)$,$ we
have the properties:%
\begin{eqnarray}
\mathcal{J}_{T^{2}}^{3} &=&0\qquad ,\qquad \mathcal{J}_{T^{2}}^{2}=0
\label{T2f} \\
\mathcal{J}_{K3}^{3} &=&0\qquad ,\qquad \mathcal{J}_{K3}^{2}\neq 0
\label{K3b}
\end{eqnarray}%
implying the vanishing of the intersection in the fiber ($J_\alpha\cdot
J_\beta=0$) and the three-intersection in the base ( $J_{a}.J_{b}.J_{c}=0)$.
To write down the gauge coupling kinetic matrix, we start by expressing the
volumes of the different cycles as:%
\begin{equation}
\mathcal{V}_{A}=\frac{1}{2}\kappa _{ABC}v^{B}v^{C}\qquad ,\qquad \mathcal{V}%
_{AB}=\frac{1}{2}\kappa _{ABC}v^{C}
\end{equation}%
where $\kappa _{ABC}$ is the three-intersection in $X_{3}$. By substituting $%
v^{A}=(\lambda ^{1/3}v^{a},\lambda ^{-2/3}\tilde{v}^{\alpha })$ where we
have inserted the dependence on the spectral parameter, we end up with:%
\begin{eqnarray}
\mathcal{V}_{a} &=&\frac{\lambda ^{-1/3}}{2}\kappa _{ab\alpha }v^{b}\tilde{v}%
^{\alpha }  \notag \\
\mathcal{\tilde{V}}_{\alpha } &=&\frac{\lambda ^{2/3}}{2}\kappa _{\alpha
bc}v^{b}v^{c}
\end{eqnarray}%
where we have used $\kappa _{abc}=\kappa _{a\alpha \beta }=0$ and $\kappa
_{\alpha \beta \gamma }=0.$ Similarly:%
\begin{eqnarray}
\mathcal{V}_{ab} &=&\frac{\lambda ^{-2/3}}{2}\kappa _{ab\gamma }\tilde{v}%
^{\gamma }  \notag \\
\mathcal{V}_{a\beta }^{\prime } &=&\frac{\lambda ^{1/3}}{2}\kappa _{a\beta
c}v^{c} \\
\mathcal{\tilde{V}}_{\alpha \beta } &=&0  \notag
\end{eqnarray}%
This leads to:%
\begin{eqnarray}
G_{ab} &=&\frac{\lambda ^{-2/3}}{2}(\frac{1}{2}\kappa _{ac\gamma }\kappa
_{bd\delta }v^{c}v^{d}\tilde{v}^{\delta }-\kappa _{ab\gamma })\tilde{v}%
^{\gamma }=\lambda ^{-2/3}\mathring{G}_{ab}  \label{coup1} \\
G_{a\alpha }^{\prime } &=&\frac{\lambda ^{1/3}}{2}(\frac{1}{2}\kappa
_{ac\gamma }\kappa _{\alpha de}v^{c}v^{e}\tilde{v}^{\gamma }-\kappa
_{a\alpha d})v^{d}=\lambda ^{1/3}\mathring{G}_{a\alpha }^{\prime }
\label{coup2} \\
\tilde{G}_{\alpha \beta } &=&\frac{\lambda ^{4/3}}{4}\kappa _{\alpha
cd}\kappa _{\beta ef}v^{c}v^{d}v^{e}v^{f}=\lambda ^{4/3}\widetilde{\mathring{%
G}}_{\alpha \beta }  \label{coup3}
\end{eqnarray}%
with $\mathring{G}_{..}$ being independent from $\lambda .$

If we restrain the curve $\mathcal{C}=q_{A}\mathcal{C}^{A}$ to: (i) purely
fibral curves $\tilde{C}=\tilde{q}_{\alpha }\tilde{C}^{\alpha }$ (i,e $%
q_{a}=0$) or (ii) pure base curves $C=q_{a}C^{a}$ (i,e $q_{\alpha }=0$)$,$
we end up with two effective field configurations with corresponding gauge
couplings $g_{\mathcal{C}}^{2}=g_{5}^{2}(q_{a}G^{ab}q_{b})$ and $\tilde{g}_{%
\mathcal{\tilde{C}}}^{2}=g_{5}^{2}(\tilde{q}_{\alpha }\tilde{G}^{\alpha
\beta }\tilde{q}_{\beta }).$ For these two particular cases we have the
following behaviours in terms of the spectral parameter as:%
\begin{eqnarray}
(i) &:&\tilde{g}_{\tilde{C}}^{2}\sim \lambda ^{-4/3}\qquad ,\qquad
g_{C}^{2}=0  \label{1} \\
(ii) &:&\text{ }g_{C}^{2}\sim \lambda ^{2/3}\qquad ,\qquad \tilde{g}_{\tilde{%
C}}^{2}=0  \label{2}
\end{eqnarray}%
on the other hand if we take a generic curve generic curve $\mathcal{C}=q_{A}%
\mathcal{C}^{A}$ which has compononets both along the fiber and the base,
the gauge coupling behaves as%
\begin{equation}
(iii):g_{\mathcal{C}}^{2}\sim \lambda ^{2/3}\mathring{G}^{ab}+\lambda ^{-1/3}%
\mathring{G}^{\prime a\alpha }+\lambda ^{-4/3}\widetilde{\mathring{G}}%
^{\alpha \beta }  \label{3}
\end{equation}%
From this result, we distinguish the following:

\textbf{case (i):} purely fibral curves:

In the limit $\lambda \rightarrow \infty ,$ the volume of the fiber shrink
according to the scaling $v^{A}=(\lambda ^{1/3}v^{a},\lambda ^{-2/3}\tilde{v}%
^{\alpha })$.\ In this case the fibral curves give a weak coupling regime.
Moreover in the F-theory picture, the shrinking of the volume of $T^{2}$ is
seen as a decompactification of the 5D theory to 6d, and towers of weakly
coupled KK states are expected to be present which are seen in the M-theory
picture as M2 branes wrapping the vanishing 2-cycles \cite{5d}.

On the other hand, by moving to the limit $\lambda \rightarrow 0$ the volume
of the fibral curves diverges, (this behaviour in the two limits has been
captured in the previous figure), as a result in this case we have a strong
coupling regime arising from the expanding fibral curves.

\textbf{case (ii):} purely base curves:

In the limit $\lambda \rightarrow \infty ,$ the volume of the base curves
diverge according to the same scaling $v^{A}=(\lambda ^{1/3}v^{a},\lambda
^{-2/3}\tilde{v}^{\alpha }),$ similarly to the previous case, diverging
curves lead to a strong coupling regime. In the other region of the moduli
space given by $\lambda \rightarrow 0$ it is the base that shrinks leading
to a weak coupling regime. The latter coincides with the emergent string
limit where the weakly coupled emergent string is given by M5 wrapping the
shrinking surface $K3$.

\textbf{case (iii):} a generic curve:

Given the generic curve $\mathcal{C}=q_{A}\mathcal{C}^{A}$, the gauge
coupling is always dominated by the strong coupling regime, this is due to
the curve having always contribution from the expanding geometry.\ I,e: in
the limit $\lambda \rightarrow \infty $ where the base expands, the gauge
coupling is dominated by $\lambda ^{2/3}\mathring{G}^{ab},$ whereas in the
limit $\lambda \rightarrow 0$ it is the fiber that expands leading to $g_{%
\mathcal{C}}^{2}\sim $ $\lambda ^{-4/3}\widetilde{\mathring{G}}^{\alpha
\beta }$ which also gives a strong coupling.

Furthermore, with the mapping $\lambda \rightarrow 1/\lambda ,$ it is
possible to relate the above regimes as it exchanges strong and weak
coupling limits between fibral and base components inducing a weak/strong
duality.\ This corresponds to threefolds of type $K3$ of the form $K3\times
T^{2}$ in (\ref{FIB}-II)$.$

In fact considering this second possible structure of the threefold: $%
X_{3}=K3\times T^{2},$\ i,e manifold of a $K3$ fibration and base $T^{2},$we
obtain the same properties as in the previous case, only with the difference
of which component is seen as the fiber and which is seen as the base:%
\begin{eqnarray}
\mathcal{J}_{K3}^{3} &=&0\qquad ,\qquad \mathcal{J}_{K3}^{2}\neq 0
\label{K3f} \\
\mathcal{J}_{T^{2}}^{3} &=&0\qquad ,\qquad \mathcal{J}_{T^{2}}^{2}=0
\label{T2b}
\end{eqnarray}%
They lead to the following vanishing intersection numbers $J_{a}\cdot
J_{b}=J_{a}.J_{b}.J_{c}=0$ for all base elements and to $J_{\alpha
}.J_{\beta }.J_{\gamma }=0$ for the fibral components. As for the volumes,
we get:%
\begin{eqnarray}
\mathcal{V}_{a} &=&\frac{1}{2}\lambda ^{-2/3}\kappa _{a\gamma \delta }\tilde{%
v}^{\gamma }\tilde{v}^{\delta }  \notag \\
\mathcal{\tilde{V}}_{\alpha } &=&\frac{1}{2}\lambda ^{1/3}\kappa _{\alpha
c\gamma }v^{c}\tilde{v}^{\gamma }
\end{eqnarray}%
and:%
\begin{eqnarray}
\mathcal{V}_{ab} &=&\frac{1}{2}\kappa _{abC}v^{C}=0  \notag \\
\mathcal{V}_{a\beta }^{\prime } &=&\frac{1}{2}\lambda ^{-1/3}\kappa _{a\beta
\gamma }\tilde{v}^{\gamma } \\
\mathcal{\tilde{V}}_{\alpha \beta } &=&\frac{1}{2}\lambda ^{2/3}\kappa
_{\alpha \beta c}v^{c}
\end{eqnarray}%
This leads to:%
\begin{eqnarray}
G_{ab} &=&\frac{\lambda ^{-4/3}}{4}\kappa _{a\gamma \delta }\kappa _{b\mu
\nu }\tilde{v}^{\gamma }\tilde{v}^{\delta }\tilde{v}^{\mu }\tilde{v}^{\nu }
\label{coup4} \\
G_{a\alpha }^{\prime } &=&\frac{\lambda ^{-1/3}}{2}(\frac{1}{2}\kappa
_{a\gamma \delta }\kappa _{\alpha c\gamma }\tilde{v}^{\gamma }\tilde{v}%
^{\delta }v^{c}-\kappa _{a\beta \gamma })\tilde{v}^{\gamma }  \label{coup5}
\\
\tilde{G}_{\alpha \beta } &=&\frac{\lambda ^{2/3}}{2}(\frac{1}{2}\kappa
_{\alpha c\gamma }\kappa _{\beta d\delta }\tilde{v}^{\gamma }v^{d}\tilde{v}%
^{\delta }-\kappa _{\alpha \beta c})v^{c}  \label{coup6}
\end{eqnarray}

Finally, we have the gauge couplings for purely fibral, base, and generic
curves:%
\begin{eqnarray}
(iv) &:&\tilde{g}_{\tilde{C}}^{2}\sim \lambda ^{-2/3}\qquad ,\qquad
g_{C}^{2}=0  \notag  \label{iv} \\
(v) &:&g_{C}^{2}\sim \lambda ^{4/3}\qquad ,\qquad \tilde{g}_{\tilde{C}}^{2}=0
\label{v} \\
(vi) &:&g_{\mathcal{C}}^{2}\sim \lambda ^{4/3}\mathring{G}^{ab}+\lambda
^{1/3}\widetilde{\mathring{G}}^{a\alpha }+\lambda ^{-2/3}\mathring{G}%
^{\prime \alpha \beta }  \notag  \label{vi}
\end{eqnarray}

These are the same as (\ref{1}-\ref{3}), with the only difference being
changing $\lambda \rightarrow 1/\lambda ,$ which shows that by comparing (%
\ref{1}-\ref{3}) to (\ref{iv}-\ref{vi}) indeed weakly coupled (resp.\
strongly coupled) directions in the limit $\lambda \rightarrow \infty $
become strongly coupled (resp.\ weakly coupled) under the mapping $\lambda
\rightarrow 1/\lambda $ and vice versa.

Furthermore, rewriting (\ref{3}) and (\ref{v}-3) and rearanging the terms we
get:%
\begin{eqnarray}
(iii) &:&g_{\mathcal{C}}^{2}\sim \lambda ^{2/3}\mathring{G}^{ab}+\lambda
^{-1/3}\mathring{G}^{\prime a\alpha }+\lambda ^{-4/3}\widetilde{\mathring{G}}%
^{\alpha \beta }  \notag \\
&:&\qquad \qquad \updownarrow \qquad \quad \updownarrow \qquad \qquad
\updownarrow  \label{map} \\
(vi) &:&g_{\mathcal{C}}^{2}\sim \lambda ^{-2/3}\widetilde{\mathring{G}}%
^{\alpha \beta }+\lambda ^{1/3}\mathring{G}^{\prime a\alpha }+\lambda ^{4/3}%
\mathring{G}^{ab}  \notag
\end{eqnarray}%
thus we see that similarly to \textbf{case (iii),} a generic curve having
all the terms (\ref{coup4}-\ref{coup6}) the mapping $\lambda \rightarrow
1/\lambda $ does indeed match the two cases where the roles of the fiber
(only greec indices in (\ref{map})) and the base (only latin indices)
exchange in terms of the contribution in the expression of the gauge
coupling. The mapping, in addition to changing the roles of curves in the
fiber and the base, also exchanges the gauge coupling regimes inherited from
the shrinking and the expanding of the corresponding volumes, with the
overall behaviour that always leads to a strong coupling limit in the
regions $\lambda \rightarrow 0,\infty $ due to the contribution from the
expanding geometry whether it is the fiber or the base. However the link
between each component independently of the other still obeys the
weak/strong gauge duality as seen in (\ref{map}) where the fiber (resp. the
base) induces a weak coupling limit in one region, and a strong coupling in
its inverse which is given by the aforementioned mapping.

As mentioned earlier, the conditions (\ref{T2f},\ref{K3b}) and (\ref{K3f},%
\ref{T2b}) are identical, we alternatively choose the fiber and the base
from the two geometries $T^{2}$\ and $K3$\ with the parameter $\lambda $\
free to move from $0$\ to $\infty $.\ Thus it would be more convenient to
opt for one geometry i,e $K3\times T^{2}$, and leave the choice of the base
and the fiber arbitrary. To summarise, we list hereafter the different fiber
and base structures and the corresponding coupling limits:

\begin{equation}
\begin{tabular}{ccc}
& $K3$ & $T^{2}$ \\ 
$\lambda \rightarrow \infty $ & shrinks$\Longrightarrow $weak coupling & 
expands$\Longrightarrow $strong coupling \\ 
$\lambda \rightarrow 0$ & expands$\Longrightarrow $strong coupling & shrinks$%
\Longrightarrow $weak coupling%
\end{tabular}%
\end{equation}

After fixing both the structure of the Calabi-Yau and the different gauge
coupling limits of our 5D EFT, we demonstrate in the next section how the
previously established results relating strong and weak couplings allows us
to go beyond the traditional perturbative analysis and explore strongly
coupled systems. This is particularly relevant from the Swampland
perspective, as unlike the other conjectures, the minimal WGC explicitly
demands not only weak gauge couplings but also the existence of strong
coupling limits. For this purpose, we investigate the spectrum of states
emerging at each regime and verify the accordance with the Swampland minimal
weak gravity conjecture.

\section{Minimal WGC on $K3\times T^{2}$}

Since its original proposal in \cite{9D}, the weak gravity conjecture has
undergone several refinements to establish a more rigorous formulation \cite%
{2E, 2F}.\ The most relevant versions of the WGC to this paper, listed in
order of importance are: The basic WGC \cite{9D}, the convex hull condition 
\cite{9E}, the tower WGC \cite{1A} and the minimal WGC \cite{min}.

In the first subsection, we briefly outline the main differences between the
different refinements leading to the minimal WGC. We show that this later
aligns with the analysis of the previous two sections, particularly in
relation to the structure of the Calabi-Yau and the weak/strong gauge
duality. In the second subsection, we examine the different towers of states
across these distinct gauge regimes and demonstrate that they indeed satisfy
the minimal WGC.

\subsection{Refinements of the WGC}

The basic version of the WGC states that in a consistent quantum theory
coupled gravity, there must exist at least one state of mass $m$ and charge $%
q$, such that the change to mass ratio satisfies the inequality%
\begin{equation}
\frac{q}{m}\geq \frac{Q}{M}|_{\mathrm{Ext}}  \label{basic}
\end{equation}%
where $Q$ and $M$ are the charge and mass of an extremal black hole. This
condition stems from the requirement that all black holes should be able to
discharge by emitting states satisfying (\ref{basic}), known as
superextremal states. The condition (\ref{basic}) can also be written as 
\cite{3H, 4A}:%
\begin{equation}
m^{2}\leq \frac{d-2}{d-3}|g_{U(1)}^{2}q^{2}M_{\mathrm{Pl,}d}^{d-2}
\label{basic2}
\end{equation}%
Per usual, $g_{U(1)}$ denotes the $U(1)$ gauge coupling, $d$ is the
spacetime dimension, and $M_{\mathrm{Pl},d}$ is the Planck mass in
d-dimensions$.$The above inequality accounts for a single U(1) gauge group.
However, in the presence of\textrm{\ }multiple U(1)s, a stronger requirement
given by the convex hull condition (CHC) \cite{9E} must hold. The CHC states
that there should exist a set of states whose charge-to-mass ratio vectors
enclose the unit ball, which coincides with the black hole region, ensuring
a stronger version of (\ref{basic}).

The tower WGC on the other hand refines both the CHC and the basic WGC.\ In
fact, the CHC does not always hold true under dimensional reduction.
Specifically, if the WGC or CHC are verified in D-dimensions, it does not
necessarily follow that they will stay satisfied in D-1 dimensions. The CHC
fails in the limit where the radius of the compactified dimension shrinks to
zero \cite{1A}. The proposed solution is to require the existence of
infinite towers of states in every direction in the charge lattice, as this
would guarantee that the CHC remains consistent under dimensional reduction
by having states of large enough charge-to-mass ratio in the infinite towers.

The tower WGC condition can be written as follows:%
\begin{equation}
M_{k}^{2}\leq \frac{d-2}{d-3}|g_{d}^{2}\vec{q}_{k}^{2}M_{\mathrm{Pl},d}^{d-2}
\label{tower}
\end{equation}%
with $M_{k}$ and $\vec{q}_{k}$ defining respectively the mass and the
quantised charge vector of the k-th state in the tower along the direction
of $\vec{q}.$ Where $g_{d}$ denotes the gauge coupling in d-dimensions$.$
This inequality must be satisfied in every direction in the charge lattice.
However, as pointed out in \cite{min}, demonstrating the existence of
infinite towers of such states in all directions with current techniques is
not always possible, which poses the question of whether the tower WGC is
the most complete version of the conjecture.

The proposed minimal WGC is a refined version of the tower WGC that resolves
all previous issues, including consistency under dimensional reduction and
the possibility of proving it in all setups. As mentioned earlier, thus far,
the most consistent version of the WGC under dimensional reduction is the
tower WGC, which requires the existence of superextremal towers of states in
every direction in the charge lattice of the theory. However the minimal WGC 
\cite{min} states that towers of superextremal states appear in a given
gauge theory coupled to gravity if and only if they are necessary for
maintaining the consistency of the WGC under dimensional reduction this was
then conjectured to fall under three categories as in:

\begin{conjecture}
Towers of superextremal states are present in a given gauge theory coupled
to gravity if and only if the theory exhibits an emergent string limit, KK
gauge bosons, or a strong coupling limit.\footnote{%
Where the second part of the assertion is taken to be equivalent to being
required by dimensional reduction.}
\end{conjecture}

The key difference from the tower version is that superextremal towers are
not always expected to exist if they are not needed. Particularly, in setups
where no towers has been found, it is because the minimal WGC does not
require them. The existence of towers is then equivalent to the presence of
an emergent string limit, KK gauge bosons, or a strong coupling limit. In
fact, the determination of whether the theory in a simple dimensional
reduction, an emergent string limit or a decompactification one for instance
is linked to the definition of the dimensional reduction. More precisely, if
the radius of the compactified circle shrinks to a limit where the KK scale
exceeds the black hole threshold, i,e the KK spectrum consists of black
holes rather than particles, then this theory can no longer be considered a
valid dimensional reduction. The occupants of the dominant tower then
defines whether we are in an emergent string limit, or a decompactification
limit.

To illustrate this, let us consider the example discussed in \cite{min} of
the circle compactification of 6D EFT arising from F-theory on an
elliptically fibered Calabi-Yau threefold. Naively, taking the limit $%
r_{S^{1}}\rightarrow 0$ yields a 5D KK theory. However in this limit$,$ the
KK scale $m_{KK}\sim \frac{1}{r_{S^{1}}}$ exceeds the black hole threshold,
meaning we are no longer in a simple 5D KK theory. Using the duality between
F-theory and M-theory on the elliptically fibered Calabi-Yau whose fiber $%
T^{2}$ has a volume $M_{\mathrm{Pl,}11}^{3}\mathcal{V}_{T^{2}}\sim 1/r_{S^{1}},$ the
compactification of M-theory on the internal manifold shows that we either
have: a 5D emergent string limit where the M5 brane wraps a shrinking
surface $\mathbb{S}$ or a decompactification limit characterised by the
divergence of the volume of an elliptic curve. This example fits nicely with
our setup where we take M-theory on $T^{2}\times \mathbb{S}$ with the
different limits arising from wrapping M2 or M5 branes on appropriate
shrinking or expanding cycles in the CY.

Subsequently, with the aim of investigating the presence of the tower and
their occupants for our $T^{2}\times K3$ model in accordance with the
minimal weak gravity conjecture, we examine the arising limits, whether
emergent string, decompactification limits strong coupling, derive the
corresponding towers of superextremal states then make connections with the
weak/strong gauge duality of the 5D EFT.

\subsection{The towers of the minimal WGC}

As discussed in the previous sections, by imposing constraints on the CY3,
we have shown that the EFT defines two main regimes: the limit $\lambda
\rightarrow 0$ leads to the emergence of KK gauge bosons, while the limit $%
\lambda \rightarrow \infty $ correspond to an emergent string. In both
cases, a strong coupling is present. The associated tower of states can be a
priori occupied by a mixture of BPS and non-BPS states.\ Following \cite{5d,
3d}, BPS states emerge from M2 branes wrapping curves of positive
self-intersection in the fiber $K3,$whereas non-BPS states are excitations
of the weakly coupled emergent string given by M5 wrapping the shrinking $%
K3. $ In this section, we will investigate the arising towers from a
Calabi-Yau of the form $K3\times T^{2}$ in both weak and strong gauge
couplings regimes and examine the type of states occupying each direction.

Among the two possible states that can inhabit the tower, BPS states are of
special interest because their charge to mass ratio is protected by
supernumerary.{\normalsize \ }This is useful for our analysis as we navigate
the moduli space by varying the parameter $\lambda $ to cover all possible
settings$.$ Specifically, a superextremal BPS state in the limit $\lambda
\rightarrow \infty $ will remain superextremal as we gradually move towards $%
\lambda \rightarrow 0$.\ Moreover, in the 5D EFT obtained by compactifying
M-theory on a Calabi-Yau threefold, it has been shown that all BPS states
arising from M2 branes wrapping movable curves do satisfy the conjecture 
\cite{9}.\ As a result, their existence alone suffices to validate this
Swampland constraint.

To proceed, we start by expressing the mass of a state arising from wrapping
an M2 brain around\ a curve $C^{A}\in H_{2}(X_{3})$ in terms of the volume
of the curve. The basis of two cycles splits as:%
\begin{equation*}
\{C^{A}\}=\{\tilde{C}^{\alpha },C^{a}\}
\end{equation*}%
A M2 brane wrapping a shrinking curve in the fiber in the limit $\lambda
\rightarrow \infty $ gives a state of mass:%
\begin{equation}
m_{\mathrm{light}}\sim vol(\tilde{C}^{\alpha })M_{\mathrm{Pl,}11}
\end{equation}%
with $vol(\tilde{C}^{\alpha })$ being the dimensionless volume of $\tilde{C}%
^{\alpha }.$ Using (\ref{T2}), we learn that the volume of the cycle $\tilde{%
C}^{\alpha }$ shrinks as $\lambda ^{-2/3}$ leading to:%
\begin{equation}
m_{\mathrm{light}}\sim \lambda ^{-2/3}M_{\mathrm{Pl,}11}
\end{equation}%
Similarly, heavy towers results from wrapping expanding curves in the base
in the same limit giving thus:%
\begin{equation*}
m_{\mathrm{heavy}}\sim \lambda ^{1/3}M_{\mathrm{Pl,}11}
\end{equation*}

We now need to verify whether these towers satisfy the conjecture.
Leveraging the fact that the WGC and the Repulsive Force Conjecture (FC)
agree in infinite distances \cite{5}, we can write the condition on the
towers of state satisfying the minimal WGC as:%
\begin{equation}
M_{k}^{2}\leq \frac{d-2}{d-3}|_{d=5}g_{5}^{2}M_{\mathrm{Pl}%
,5}^{3}((kq_{A})G^{AB}(kq_{B}))-3(G^{AB}-\frac{1}{3}\hat{v}^{A}\hat{v}^{B})%
\frac{\partial M_{k}}{\partial \hat{v}^{A}}\frac{\partial M_{k}}{\partial 
\hat{v}^{B}}  \label{ineq}
\end{equation}%
Where the $\hat{v}^{A}=\frac{v^{A}}{\mathcal{V}^{1/3}}$ are the rescaled
Kahler moduli and $M_{k}$ is the mass of a state at level $k$ in the tower,
it is proportional to the mass scale $m_{\mathrm{light}}$ or $m_{\mathrm{%
heavy}}$ depending on the tower. The integer $k$ is also related to the
charge of the states since it captures the number of times the M2 brane
wraps a given curve, this define the first term of the right hand side (RHS) of (\ref{ineq}).
This inequality aligns with the Repulsive Force Condition (RFC) which states
that in a consistent theory of gravity, attractive forces $F_{\mathrm{%
Attractive}}=F_{\mathrm{Gravity}}+F_{\mathrm{Yukawa}}$ should not dominate
over repulsive forces $F_{\mathrm{Coulomb}}=F_{\mathrm{repulsive}}$:%
\begin{equation*}
F_{\mathrm{Gravity}}+F_{\mathrm{Yukawa}}\leq F_{\mathrm{Coulomb}}
\end{equation*}%
Where the Yukawa term in our setting is given by: $G^{AB}=\frac{1}{2}%
\mathfrak{g}^{XY}\partial _{X}\hat{v}^{A}\partial _{Y}\hat{v}^{B}+\frac{1}{3}%
\hat{v}^{A}\hat{v}^{B}$ where $\mathfrak{g}^{XY}$ is the inverse of the
Yukawa coupling matrix (see Appendix). In the absence of the Yukawa force,
this provides an intuitive motivation for the Weak Gravity Conjecture
demanding that gravity must be the weakest force as one then recovers the
regular expression of the tower WGC (\ref{tower}).

Let us consider M2 branes wrapping $k$-times curves of the form: $C_{f}=\sum 
\tilde{q}_{\alpha }\tilde{C}^{\alpha }.$ The corresponding mass of the
arising BPS state is:%
\begin{equation}
M_{k}=2\pi k\tilde{q}_{\alpha }\tilde{v}^{\alpha }M_{\mathrm{Pl,}11}=2\pi k%
\tilde{q}_{\alpha }\widehat{\tilde{v}}^{\alpha }\frac{M_{\mathrm{Pl,}5}}{%
(4\pi )^{1/3}}
\end{equation}%
thus giving:%
\begin{equation}
M_{k}^{2}=4\pi ^{2}k^{2}\tilde{q}_{\alpha }\tilde{q}_{\beta }\frac{\widehat{%
\tilde{v}}^{\alpha }\widehat{\tilde{v}}^{\beta }}{(4\pi )^{2/3}}M_{\mathrm{Pl%
},5}^{2}  \label{mk}
\end{equation}%
To verify the validity of the (\ref{ineq}), we explicitly write the first
term of the RHS:%
\begin{eqnarray}
\frac{d-2}{d-3}|_{d=5}g_{5}^{2}M_{\mathrm{Pl},5}^{3}((k\tilde{q}_{\alpha })%
\tilde{G}^{\alpha \beta }(k\tilde{q}_{\beta })) &=&\frac{3}{2}(2\pi )(4\pi
)^{1/3}k^{2}M_{\mathrm{Pl},5}^{2}(\frac{1}{2}\tilde{q}_{\alpha }\tilde{v}%
^{\alpha }\tilde{v}^{\beta }\tilde{q}_{\beta }-\tilde{q}_{\alpha }\widehat{%
\mathcal{\tilde{V}}}^{\alpha \beta }\tilde{q}_{\beta })  \notag \\
&=&3\times 4^{1/3}\times \pi ^{4/3}k^{2}M_{\mathrm{Pl},5}^{2}(\frac{1}{2}%
\tilde{q}_{\alpha }\tilde{v}^{\alpha }\tilde{v}^{\beta }\tilde{q}_{\beta }-%
\tilde{q}_{\alpha }\widehat{\mathcal{\tilde{V}}}^{\alpha \beta }\tilde{q}%
_{\beta })
\end{eqnarray}%
while the second term reads:%
\begin{eqnarray}
3(\tilde{G}^{\alpha \beta }-\frac{1}{3}\widehat{\tilde{v}}^{\alpha }\widehat{%
\tilde{v}}^{\beta })\frac{\partial M_{k}}{\partial \widehat{\tilde{v}}%
^{\alpha }}\frac{\partial M_{k}}{\partial \widehat{\tilde{v}}^{\beta }} &=&3(%
\frac{1}{6}\widehat{\tilde{v}}^{\alpha }\widehat{\tilde{v}}^{\beta }-%
\widehat{\mathcal{\tilde{V}}}^{\alpha \beta })\frac{(2\pi k)^{2}}{(4\pi
)^{2/3}}\tilde{q}_{\alpha }\tilde{q}_{\beta }  \notag \\
&=&3\times 4^{1/3}\times \pi ^{4/3}k^{2}M_{Pl,5}^{2}(\frac{1}{6}\tilde{q}%
_{\alpha }\tilde{v}^{\alpha }\tilde{v}^{\beta }\tilde{q}_{\beta }-\tilde{q}%
_{\alpha }\widehat{\mathcal{\tilde{V}}}^{\alpha \beta }\tilde{q}_{\beta })
\end{eqnarray}%
Subtracting the 2 terms, we get for the RHS:%
\begin{equation}
RHS=4^{1/3}\times \pi ^{4/3}k^{2}M_{\mathrm{Pl},5}^{2}\tilde{q}_{\alpha }%
\widehat{\tilde{v}}^{\alpha }\widehat{\tilde{v}}^{\beta }\tilde{q}_{\beta }
\end{equation}%
corresponding exactly to (\ref{mk}). Therefore, tower of states arising from
M2-branes wrapping 2-cycles all saturates (\ref{ineq}). These towers can be
classified based on their mass. By considering shrinking and expanding
cycles, we obtain light, weakly coupled towers and heavy, strongly coupled
towers as follows:

\begin{enumerate}
\item Light, weakly coupled towers:
\end{enumerate}

\begin{itemize}
\item The first type of towers is populated by \textit{light} BPS states
denoted $\mathcal{T}_{M_{k\rightarrow 0}}^{T^{2}}$. It arises for $\lambda
\rightarrow 0$ where the torus shrinks leading to a type-$T^{2}$ limit. The
shrunken curves $C^{\alpha }$ give rise to BPS states realised by M2 branes
wrapping the 2-cycles of the basis of fibral curves.\ Since the mass of
these BPS states is proportional to the shrinking volume, the resulting
tower of BPS states is light. These states correspond to the weakly coupled
mentioned in \cite{5d}.

\item A second tower of light states emerges in the limit $\lambda
\rightarrow \infty ,$ where the volume of the torus $T^{2}$ diverges. This
is equivalent to considering a type $K3$ limit in \cite{5d} with a shrinking
surface.\ Recall that the twofold $K3$ has a lattice $\Gamma ^{3,19}$ and
contains curves with positive self-intersection. Provided that the $K3$
surface does not fully degenerate, or that it only degenerates at finite
distance \cite{5d}, we get a tower of \textit{light} weakly coupled BPS
states $\mathcal{T}_{M_{k\rightarrow 0}}^{K3}$, realised by M2 branes
wrapping shrinking curves in the K3 surface.
\end{itemize}

\begin{enumerate}
\item[2.] Heavy strongly coupled towers:
\end{enumerate}

\begin{itemize}
\item The \textit{heavy} tower $\mathcal{T}_{M_{k\rightarrow \infty }}^{K3}$%
consists of the same M2 branes wrapping\textrm{\ }the same curves with
positive self-intersection in the surface K3 as in the previous case.
However, instead of staying in the limit $\lambda \rightarrow \infty ,$ we
now continuously move $\lambda $ throughout the moduli space from $\lambda
\rightarrow \infty $ to $\lambda \rightarrow 0$. In consequence, the
previously light towers now become heavy seeing that the K3 surface expands
in this limit. Nevertheless, the tower remains superextremal due to its
charge to mass ratio being protected by supersymmetry.

\item Similarly, another tower of BPS states arises from $T^{2}$, and we
label them as $\mathcal{T}_{M_{k\rightarrow \infty }}^{T^{2}}$ representing
towers of heavy strongly coupled BPS states formed by wrapping curves in the
expanding $T^{2}$ in the limit $\lambda \rightarrow \infty $.
\end{itemize}

As expected from the minimal WGC\textrm{\ }and the link between weak and
strong couplings, we first find that towers of weakly and strongly coupled
(light and heavy) states satisfy the WGC. All these towers consist of BPS
states, obtained by wrapping M2 branes on some holomorphic movable curves,
either on the fiber or the base. The specific choice of type $T^{2}$ or type 
$K3$ limit determines which curves are wrapped.

It is worth noting that towers of non-BPS states are also expected to appear
in the theory.\ As shown in \cite{5d}, for threefolds of type $K3$ of the
form $K3\times \mathbb{P}^{1},$ towers of weakly coupled non-BPS states
exist and they satisfy the conjecture. For our case, such towers are also
expected to be superextremal. Note also that while the BPS states arising
from $T^{4}$ result from M2 branes wrapping curves with positive self
intersection, the non-BPS states correspond to excitations of the string
formed by an M5 brane wrapping $T^{4}$.

\section{Conclusion and comments}

In this paper, we investigated aspects of the minimal WGC in the presence of
towers of both strongly and weakly coupled states. Our study targeted 5D
EFTs arising from M-Theory compactifications on a Calabi-Yau threefold with
finite volume. The novelty of our approach lies in allowing both the fiber
and the base to shrink or expand provided that the volume remains finite.
This constraint naturally led us to consider threefolds of the form $%
X_{3}=K3\times T^{2}$\ where the mapping $\lambda \rightarrow 1/\lambda $
exchanges the shrinking with\ the expanding entity i,e fiber and base. This
extends the results of the fourfold in the more obvious case of $%
X_{4}=K3\times K3$ \cite{3d}.

The aforementioned correspondence represented by the mapping $\lambda
\rightarrow 1/\lambda $ which exchanges different cycles in the manifold,
also exchanges the strong and weak coupling limits.\ After defining the
different gauge regimes, it remains to prove the existence of towers of
superextremal states.\ We identified four towers of BPS states that arise
from M2 branes wrapping holomorphic movable curves in the manifold. As
expected from the minimal WGC, two of these towers correspond to heavy,
strongly coupled states $\mathcal{T}_{M_{k\rightarrow \infty }}^{T^{2}}$ and 
$\mathcal{T}_{M_{k\rightarrow \infty }}^{K3}$, while the other two define
light, weakly coupled states $\mathcal{T}_{M_{k\rightarrow 0}}^{T^{2}}$ and $%
\mathcal{T}_{M_{k\rightarrow 0}}^{K3}$ which appears in the emergent string
limit or as duals to KK gauge bosons.

The correspondence discussed in this paper can also be linked to the
distance conjecture \cite{11, 12}, which states that along an infinite
geodesic distance in the moduli space, a tower of states becomes
asymptotically massless.\ Note however that in many cases an intriguing
pattern often occurs: the asymptotically massless tower of states is always
accompanied by an asymptotically heavy tower.\ One possible explanation was
mentioned in \cite{13}, is that masses are parameterised by the expectation
values of scalar fields. Consequently, when a mass scale becomes
super-Planckian i,e $m_{i}\gg M_{\mathrm{Pl},d}$, a scalar field becomes
subject to significant growth, bringing the distance conjecture into effect.
A well-known example illustrating this phenomenon is string theory on a
circle. In this instance, wrapping a string on a circle gives rise to two
towers of states: one corresponding to winding modes and the other to
Kaluza-Klein modes whose masses are inversely proportional to each other.

This exact pattern also occurs in the threefold of our model. Specifically,
when taking either $\lambda \rightarrow 0$ or $\lambda \rightarrow \infty ,$
two accompanying towers of states always emerge, one heavy and one light,
such that their scaling compensates each other. In fact according to the
distance conjecture, the mass scale of cheese towers are of the form $m\sim
e^{-\alpha \Delta \phi }$ with $|\Delta \phi |$ being a geodesic distance in
the moduli space that becomes infinite i,e $\Delta \phi \rightarrow \pm
\infty .$ We can make connection with this exponential behaviour by
considering $\log \lambda $ which clearly shows that as $\lambda \rightarrow
0$, we obtain $\log \lambda \rightarrow -\infty ,$ while for $\lambda
\rightarrow \infty $ we have $\log \lambda \rightarrow \infty .$ This
confirms that $\lambda $ parameterises the geodesic distance in the moduli
space.

Finally, the strong/weak gauge duality bears a resemblance to T-duality,
which was first introduced for fourfolds of the form $K3\times K3$, using a
similar rationale to the one presented in this paper.\ This idea could
potentially be extended to fourfolds of the form $T^{2}\times CY3$ and to
the simpler twofold case $T^{4}=T^{2}\times T^{2}.\ $And while in our
analysis this duality was primarily used as a test of the minimal WGC, it
may also have deeper implications for the EFT, warranting further
investigation.

\section{Appendix: 5D EFT from M-theory on $K3\times T^{2}$}

In this Appendix, We briefly review the compactification of M-theory on
Calabi-Yau threefolds in preparation to relate the fibration structure to
the different gauge coupling limits.

We start with the analysis of the low energy effective action of 11d
supergravity which, in the notation of \cite{5d}, takes the following form:%
\begin{equation}
\mathcal{S}_{11D}=2\pi M_{\mathrm{Pl},11}^{9}\int_{\mathcal{M}_{11d}}(%
\mathcal{R}_{11}\ast \mathbf{1}-\frac{1}{2}\boldsymbol{F}_{4}\wedge \ast (%
\boldsymbol{F}_{4}))+...
\end{equation}%
Where $M_{11}$ is the 11d Planck mass, $\mathcal{R}_{11}$ is the 11d Ricci
tensor, and $F_{4}$ is the field strength of the gauge potential $C_{3}$.
After compactification on a general Calabi-Yau threefold, the 5D effective
action reads as:%
\begin{equation}
\mathcal{S}_{5D}=\frac{M_{\mathrm{Pl},5}^{3}}{2}\int_{\mathcal{M}%
_{5D}}\mathcal{R}_{11}\ast \mathbf{1-}\mathfrak{g}_{XY}d\Phi ^{X}\wedge \ast
d\Phi ^{Y}-\frac{1}{2g_{5}^{2}}\int_{\mathcal{M}_{5D}}G_{AB}\boldsymbol{F}%
^{A}\wedge \ast \boldsymbol{F}^{B}
\end{equation}%
with $\mathfrak{g}_{XY}$ the Yukawa coupling matrix. Under compactification,
the gauge filed $\boldsymbol{C}_{3}$ splits as:%
\begin{equation}
C_{3}=(2\pi )^{-1}M_{\mathrm{Pl},11}^{-1}A^{A}\wedge J_{A}  \label{R1}
\end{equation}%
where the $J_{A},$ $A=1,...,h^{1,1}(X_{3})$ form a basis of Kahler form
generators%
\begin{equation}
J=v^{A}J_{A}  \label{R2}
\end{equation}%
where $v^{A}$ \footnote{%
Note that the scaling $\lambda $ is included in $v^{A}$ for lighter notations%
} is the dimensionless volume of the 2-cycle $\mathcal{C}^{A}.$ the two
relations (\ref{R1}, \ref{R2}) define a duality between the elements of $%
H^{1,1}(X_{3},\mathbb{Z})$ and a basis of gauge groups denoted as $%
\{U(1)^{A}\}.$

Moreover, the scalar fields $\Phi ^{X},$ $X=1,...,h^{1,1}(X_{3})-1$ generate
the Yukawa coupling which is crucial for the WGC, since at infinite
distances in the moduli space the WGC becomes equivalent to the Repulsive
force condition as previously mentioned in \cite{5d, 10}.

The constants in the 5D effective action are given by:%
\begin{equation}
M_{\mathrm{Pl},5}^{3}=4\pi M_{\mathrm{Pl},11}^{3}\mathcal{V\qquad },\qquad
g_{5}^{2}=\frac{2\pi (4\pi )^{1/3}}{M_{\mathrm{Pl},5}}
\end{equation}%
And the coupling matrix $G_{AB}$ reads as$:$%
\begin{eqnarray}
G_{AB} &=&\frac{1}{\mathcal{V}^{1/3}}\int_{X_{3}}J_{A}\wedge \ast J_{B} \\
&=&\mathcal{\hat{V}}_{A}\mathcal{\hat{V}}_{B}\mathcal{-\hat{V}}_{AB}
\end{eqnarray}%
with%
\begin{eqnarray}
\mathcal{\hat{V}}_{A} &=&\frac{1}{\mathcal{\hat{V}}^{2/3}}\mathcal{V}_{A}=%
\frac{1}{2\mathcal{\hat{V}}^{2/3}}\int_{X_{3}}J_{A}\wedge J^{2}=\frac{1}{2}%
\kappa _{ABC}v^{B}v^{C} \\
\mathcal{\hat{V}}_{AB} &=&\frac{1}{\mathcal{\hat{V}}^{1/3}}\mathcal{V}_{AB}=%
\frac{1}{\mathcal{\hat{V}}^{1/3}}\int_{X_{3}}J_{A}\wedge J_{B}\wedge J=\frac{%
1}{2}\kappa _{ABC}v^{C}
\end{eqnarray}%
and:%
\begin{equation}
\kappa _{ABC}=\int_{X_{3}}J_{A}\wedge J_{B}\wedge J_{C}
\end{equation}%
Finally the gauge kinetic matrix is related to the Yukawa coupling matrix
via:%
\begin{equation}
G^{AB}=\frac{1}{2}\mathfrak{g}^{XY}\partial _{X}\hat{v}^{A}\partial _{Y}\hat{%
v}^{B}+\frac{1}{3}\hat{v}^{A}\hat{v}^{B}  \label{Yuk}
\end{equation}%
with:%
\begin{equation}
\hat{v}^{A}=\frac{v^{A}}{\mathcal{V}^{1/3}}
\end{equation}%
The associated inverse matrix is given by:%
\begin{equation}
G^{AB}=\frac{1}{2}\hat{v}^{A}\hat{v}^{B}-\mathcal{\hat{V}}^{AB}
\end{equation}%
Notice that for a curve of the form:%
\begin{equation}
\mathcal{C=}c_{A}C^{A}
\end{equation}%
the Young-Mills coupling for such a curve is given by:%
\begin{equation*}
g_{YM,\mathcal{C}}^{2}=g_{5}^{2}(c_{A}G^{AB}c_{B})
\end{equation*}%
this exhibits the weak and strong coupling limits in terms of the curves of
the internal manifold.and is used in section 3 to investigate the
weak/strong gauge duality leading to testing the minimal WGC.

\begin{equation*}
\end{equation*}


\begin{thebibliography}{99}
\bibitem{1A} Andriolo, S., Junghans, D., Noumi, T., \& Shiu, G. (2018). A
tower weak gravity conjecture from infrared consistency. Fortschritte der
Physik, 66(5), 1800020.

\bibitem{1B} Heidenreich, B., Reece, M., \& Rudelius, T. (2017). Evidence
for a sublattice weak gravity conjecture. Journal of High Energy Physics,
2017(8), 1-40.

\bibitem{2A} Arkani-Hamed, N., Motl, L., Nicolis, A., \& Vafa, C. (2007).
The string landscape, black holes and gravity as the weakest force. Journal
of High Energy Physics, 2007(06), 060.

\bibitem{2B} Montero, M., Shiu, G., \& Soler, P. (2016). The weak gravity
conjecture in three dimensions. Journal of High Energy Physics, 2016(10),
1-36.

\bibitem{2C} Sammani, R., \& Saidi, E. H. (2024). Higher spin swampland
conjecture for massive AdS \$ \_ \{3\} \$ gravity. arXiv preprint
arXiv:2406.09151.

\bibitem{2E} Harlow, D., Heidenreich, B., Reece, M., \& Rudelius, T. (2023).
Weak gravity conjecture. Reviews of Modern Physics, 95(3), 035003.

\bibitem{2F} Rudelius, T. (2024). An Introduction to the Weak Gravity
Conjecture. Contemporary Physics, 1-14.

\bibitem{3A} Vafa, C. (2005). The String landscape and the swampland. arXiv
preprint hep-th/0509212.

\bibitem{3B} Sammani, R., \& Saidi, E. H. (2024). Higher spin swampland
conjecture for massive AdS \$ \_ \{3\} \$ gravity. arXiv preprint
arXiv:2406.09151.

\bibitem{3D} Sammani, R., Boujakhrout, Y., Laamara, R. A., \& Drissi, L. B.
(2024). Finiteness of 3D higher spin gravity Landscape. Classical and
Quantum Gravity, 41(21), 215012.

\bibitem{3E} Agmon, N. B., Bedroya, A., Kang, M. J., \& Vafa, C. (2022).
Lectures on the string landscape and the Swampland. arXiv preprint
arXiv:2212.06187.

\bibitem{3F} van Beest, M., Calder\'{o}n-Infante, J., Mirfendereski, D., \&
Valenzuela, I. (2022). Lectures on the swampland program in string
compactifications. Physics Reports, 989, 1-50.

\bibitem{3H} Palti, E. (2019). The swampland: introduction and review.
Fortschritte der Physik, 67(6), 1900037.


\bibitem{4A} Heidenreich, B., Reece, M., \& Rudelius, T. (2016). Sharpening
the weak gravity conjecture with dimensional reduction. Journal of High
Energy Physics, 2016(2), 1-41.

\bibitem{5} Lee, S. J., Lerche, W., \& Weigand, T. (2019). A stringy test of
the scalar weak gravity conjecture. Nuclear Physics B, 938, 321-350.

\bibitem{6} Lee, S. J., Lerche, W., \& Weigand, T. (2019). Modular fluxes,
elliptic genera, and weak gravity conjectures in four dimensions. Journal of
High Energy Physics, 2019(8), 1-78.

\bibitem{7} Klaewer, D., Lee, S. J., Weigand, T., \& Wiesner, M. (2021).
Quantum corrections in 4d N= 1 infinite distance limits and the weak gravity
conjecture. Journal of High Energy Physics, 2021(3), 1-82.

\bibitem{5d} Cota, C. F., Mininno, A., Weigand, T., \& Wiesner, M. (2023).
The asymptotic weak gravity conjecture in M-theory. Journal of High Energy
Physics, 2023(8), 1-49.

\bibitem{3d} Charkaoui, M., Sammani, R., Saidi, E. H., \& Laamara, R. A.
(2024). Asymptotic Weak Gravity Conjecture in M-theory on $K3\times \ K3$.
Progress of Theoretical and Experimental Physics, 2024(7).

\bibitem{min} Cota, C. F., Mininno, A., Weigand, T., \& Wiesner, M. (2024).
The minimal weak gravity conjecture. Journal of High Energy Physics,
2024(5), 1-59.

\bibitem{8} Lee, S. J., Lerche, W., \& Weigand, T. (2022). Emergent strings
from infinite distance limits. Journal of High Energy Physics, 2022(2),
1-105.

\bibitem{9} Alim, M., Heidenreich, B., \& Rudelius, T. (2021). The weak
gravity conjecture and BPS particles. Fortschritte der Physik, 69(11-12),
2100125.

\bibitem{9A} Huang, Y. C., \& Taylor, W. (2019). Mirror symmetry and
elliptic Calabi-Yau manifolds. Journal of High Energy Physics, 2019(4), 1-31.

\bibitem{9B} Katz, S., Klemm, A., \& Vafa, C. (1999). M-theory, topological
strings and spinning black holes. arXiv preprint hep-th/9910181.

\bibitem{9C} Lambert, N. (2008). The M5-brane on K3$\times $T2. Journal of
High Energy Physics, 2008(02), 060.

\bibitem{9D} Arkani-Hamed, N., Motl, L., Nicolis, A., \& Vafa, C. (2007).
The String landscape, black holes and gravity as the weakest force. Journal
of High Energy Physics, 2007(06), 060.

\bibitem{9E} Cheung, C., \& Remmen, G. N. (2014). Naturalness and the weak
gravity conjecture. Physical review letters, 113(5), 051601.

\bibitem{10} Heidenreich, B., Reece, M., \& Rudelius, T. (2019). Repulsive
forces and the weak gravity conjecture. Journal of High Energy Physics,
2019(10), 1-50.

\bibitem{11} Ooguri, H., \& Vafa, C. (2007). On the Geometry of the String
Landscape and the Swampland. Nuclear physics B, 766(1-3), 21-33.

\bibitem{12} Castellano, A., Ruiz, I., \& Valenzuela, I. (2023). Stringy
evidence for a universal pattern at infinite distance. arXiv preprint
arXiv:2311.01536.

\bibitem{13} Rudelius, T. (2023). Revisiting the refined distance
conjecture. Journal of High Energy Physics, 2023(9), 1-18.
\end{thebibliography}
\end{document}